\documentclass[aps,prl,reprint,superscriptaddress]{revtex4-1}
\usepackage[english]{babel}
\usepackage[utf8]{inputenc}
\usepackage[T1]{fontenc}
\usepackage{amsmath,amssymb,braket,slashed,mathrsfs}
\usepackage[squaren,Gray,cdot,binary]{SIunits}
\usepackage{graphicx}
\graphicspath{{./figures/}}
\usepackage{color,hyperref}
\usepackage{cleveref}
\newcommand{\crefrangeconjunction}{--}
\crefformat{section}{Sec. #2#1#3}
\crefmultiformat{section}{Sec.~#2#1#3}
    { and~#2#1#3}{, #2#1#3}{ and~#2#1#3}
\crefrangeformat{section}{Sec.~#3#1#4\,\crefrangeconjunction\,#5#2#6}
\crefformat{subsection}{Sec. #2#1#3}
\crefmultiformat{subsection}{Sec.~#2#1#3}
    { and~#2#1#3}{, #2#1#3}{ and~#2#1#3}
\crefrangeformat{subsection}{Sec.~#3#1#4\,\crefrangeconjunction\,#5#2#6}
\crefformat{figure}{Fig. #2#1#3}
\crefmultiformat{figure}{Figs.~#2#1#3}
    { and~#2#1#3}{, #2#1#3}{ and~#2#1#3}
\crefrangeformat{figure}{Figs.~#3#1#4\,\crefrangeconjunction\,#5#2#6}
\crefformat{table}{Table #2#1#3}
\crefmultiformat{table}{Tabs.~#2#1#3}
    { and~#2#1#3}{, #2#1#3}{ and~#2#1#3}
\crefrangeformat{table}{Tabs.~#3#1#4\,\crefrangeconjunction\,#5#2#6}
\crefformat{equation}{(#2#1#3)}
\crefmultiformat{equation}{(#2#1#3)}{ and~(#2#1#3)}{, (#2#1#3)}{ and~(#2#1#3)}
\crefrangeformat{equation}{(#3#1#4)\crefrangeconjunction(#5#2#6)}

\definecolor{linkcolor}{rgb}{.17578125,.1875,.5703125}
\hypersetup{
    pdfstartview={FitH},    
    pdftitle={Up and down quark masses and corrections to 
    Dashen’s theorem from lattice QCD with quenched QED},    
    pdfauthor={Budapest-Marseille-Wuppertal collaboration},           
    pdfsubject={Physical Review Letters},                
    pdfkeywords={particle} {physics} {quantum} {chromodynamics} {electrodynamics} {QCD} {QED} {isospin} {hadrons} {baryons} {lattice},   
    colorlinks=true,        
    linkcolor=linkcolor,    
    citecolor=linkcolor,    
    filecolor=black,        
    urlcolor=linkcolor
}

\newcommand{\cf}{\textit{cf.}~}

\renewcommand{\bar}{\overline}
\DeclareMathOperator{\bigo}{O}

\newcommand{\dm}{\delta m}
\newcommand{\DM}{\Delta M}
\newcommand{\DQCDM}{\Delta_{\QCD}M}
\newcommand{\DQEDM}{\Delta_{\QED}M}
\renewcommand{\epsilon}{\varepsilon}
\newcommand{\QCD}{\mathrm{QCD}}
\newcommand{\QED}{\mathrm{QED}}
\DeclareMathOperator{\SU}{SU}
\newcommand{\eV}{\electronvolt}
\newcommand{\fm}{\femto\meter}
\newcommand{\budapest}{\affiliation{Inst.\ for Theor.\ Physics, E\"otv\"os
    University, P\'azm\'any P. s\'et.\ 1/A, H-1117 Budapest, Hungary}}
\newcommand{\juelich}{\affiliation{IAS/JSC, Forschungszentrum J\"ulich, D-52425
    J\"ulich, Germany}}
\newcommand{\marseille}{\affiliation{Centre de Physique Théorique, CNRS /
    Aix-Marseille U. / U. de Toulon (UMR 7332), Case
    907, F-13288 Marseille CEDEX 9, France}}
\newcommand{\wuppertal}{\affiliation{Department of Physics, Wuppertal
    University, Gaussstr. 20, D-42119 Wuppertal, Germany}}
\newcommand{\prlsec}[1]{\medskip\noindent{\em #1.}}

\begin{document}

    \title{Up and down quark masses and corrections to Dashen’s theorem\\ from 
    lattice QCD and quenched QED}
    \author{Z.~Fodor}
    \wuppertal
    \budapest
    \juelich
    \author{C.~Hoelbling}
    \wuppertal
    \author{S.~Krieg}
    \wuppertal
    \juelich
    \author{L.~Lellouch}
    \marseille
    \author{Th.~Lippert}
    \juelich
    \author{A.~Portelli}
    \marseille
    \affiliation{School of Physics \&\ Astronomy, University of Southampton, 
    SO17 1BJ, UK}
 	  \affiliation{School of Physics \&\
     Astronomy, The University of Edinburgh, EH9 3FD, UK}
    \author{A.~Sastre}
    \wuppertal
    \marseille
    \author{K.K.~Szabo}
    \wuppertal
    \juelich
    \author{L.~Varnhorst}
    \wuppertal
    
    \collaboration{Budapest-Marseille-Wuppertal collaboration}

    \date{\today}

    \pacs{PACS}

    \begin{abstract}
    In a previous letter \citep{Borsanyi:2013lga} we determined the isospin
    mass splittings of the baryon octet from a lattice calculation based on
    quenched QED and $N_f{=}2{+}1$ QCD simulations with 5 lattice spacings down
    to $\unit{0.054}{\fm}$, lattice sizes up to $\unit{6}{\fm}$ and average
    up-down quark masses all the way down to their physical value. Using the
    same data we determine here the corrections to Dashen's theorem and the
    individual up and down quark masses. For the parameter which quantifies
    violations to Dashens's theorem, we obtain $\epsilon=0.73(2)(5)(17)$, where
    the first error is statistical, the second is systematic, and the third is
    an estimate of the QED quenching error. For the light quark masses we
    obtain, $m_u=\unit{2.27(6)(5)(4)}{\mega\eV}$ and
    $m_d=\unit{4.67(6)(5)(4)}{\mega\eV}$ in the $\bar{\mathrm{MS}}$ scheme at
    $\unit{2}{\giga\eV}$ and the isospin breaking ratios
    $m_u/m_d=0.485(11)(8)(14)$, $R=38.2(1.1)(0.8)(1.4)$ and
    $Q=23.4(0.4)(0.3)(0.4)$. Our results exclude the $m_u=0$ solution to the
    strong CP problem by more than $24$ standard deviations.
    \end{abstract}

    \maketitle

    The up ($u$) and down ($d$) quark masses are two fundamental parameters of
    the Standard Model of Particle Physics. These masses cannot be directly
    determined through experiment because of the confinement of quarks within
    hadrons. Lattice QCD provides an \emph{ab-initio} approach to the
    non-perturbative calculation of QCD correlation functions. This method can
    be used to determine the light quark masses from the experimental values of
    hadron masses. In earlier work \citep{Durr:2010aw,Durr:2010vn}, we
    determined precisely $m_{ud}$, the average of the up and down quark masses,
    using lattice QCD simulations at the physical values of the quark masses.
    This quantity has also been studied by many other lattice collaborations
    (\cf the FLAG review \citep{Aoki:2013ldr}) and considerable progress has
    been made on its determination. Thus, it is now relevant to aim for the
    calculation of the light-quark mass difference $\dm=m_u-m_d$. This quantity
    is more difficult to obtain than $m_{ud}$. Its small effect on hadron
    masses, of order $\bigo(\dm/\Lambda_{\QCD})\simeq\bigo(1\%)$, is expected
    to be comparable in size to the leading $\bigo(\alpha)$ electromagnetic
    (EM) corrections, usually not included in lattice simulations. Thus,
    earlier lattice calculations of this mass difference
    \citep{Aubin:2004fs,Bazavov:2009fk,Durr:2010vn,Durr:2010aw} relied on
    phenomenological estimates of EM corrections. The inclusion of quenched QED
    effects was first performed in \citep{Duncan:1996xy} on quenched QCD
    configurations, in \citep{Blum:2007cy,deDivitiis:2013xla} on $N_f=2$ QCD
    configurations and in \citep{Blum:2010ym} on $N_f=2+1$ configurations, at a
    single lattice spacing, with rather large pion masses and in small volumes.
    Preliminary results for the present calculation can be found in
    \citep{Portelli:2010yn,Lellouch:confX,Portelli:2015wna} and preliminary
    results by MILC, in \citep{Basak:2015lla}. Very recently, an $N_f=2+1$
    calculation of $m_u/m_d$ in which QED effects are unquenched was presented
    in \citep{Horsley:2015vla}. This calculation is performed on three
    ensembles at a single lattice spacing, in volumes up to
    $(\unit{3.3}{\fm})^3$ and with sea quark masses fixed at the $\SU(3)$
    symmetric point $M_{\pi}=M_{K}\simeq\unit{412}{\mega\eV}$.
    
    Here we include QED effects to the dynamics of the valence quarks, atop
    $N_f=2+1$ QCD configurations generated directly at the physical value of
    the light-quark masses, with full continuum and infinite-volume
    extrapolations, as well as with full non-perturbative renormalization and
    running. This is a sequel to the letter \citep{Borsanyi:2013lga} which uses
    the same data set to compute light octet baryon isospin mass splittings.
    Note that a fully unquenched calculation of octet baryon and other hadron
    mass isospin splittings, with pion masses down to $\unit{195}{\mega\eV}$,
    can now be found in \citep{Borsanyi:2014jba}. Here, because we are dealing
    with light quark masses whose extraction requires reaching deep into the
    chiral regime \citep{Durr:2013goa}, we favor the simulations used in
    \citep{Borsanyi:2013lga}. This data set also has the notable advantage that
    it has been used to determine the $s$ and average $u$-$d$ quark masses in
    \citep{Durr:2010vn,Durr:2010aw}. Thus, all of the relevant non-perturbative
    renormalization and running has already been performed in pure QCD
    \citep{Durr:2010aw}.
    
    The light quark mass difference $\dm$ is connected, through a low energy
    theorem \citep{Gasser:1984pr}, to the pseudoscalar meson EM mass
    splittings. In the late 1960's, Dashen showed that pions and kaons receive
    the same EM contributions in the $\SU(3)$ chiral limit
    \citep{Dashen:1969eg}. This result is commonly known as \emph{Dashen's
    theorem}. During the 1990's, attempts to compute the chiral corrections to
    Dashen's theorem in effective field theories led to controversial and
    surprisingly large results (\cf the review \citep{Portelli:2015wna} for
    more details). In this letter we present a computation of these corrections
    from our lattice QCD and quenched QED simulations.
    
    \prlsec{General strategy}
    We consider in this work only the leading $\bigo(\alpha,\dm)$
    corrections to isospin symmetry. As was done in
    \citep{Borsanyi:2013lga}, we define $\DM^2$ to be the difference of the
    squared masses of the ``connected'' $\bar{u}u$ and $\bar{d}{d}$
    pseudoscalar mesons. It is known from partially-quenched chiral perturbation
    theory coupled to photons (PQ$\chi$PT+QED) \citep{Bijnens:2006mk} that this
    quantity is related to $\dm$ by the following expansion:
    \begin{equation}
        \DM^2=2B_2\dm+\bigo(m_{ud}\alpha,m_{ud}\dm,\alpha^2,\alpha\dm,\dm^2)
    \end{equation}
    where $B_2$ is the two-flavor chiral condensate parameter. If the quark
    masses have their physical values, we can safely make the assumption that
    $\bigo(m_{ud})=\bigo(\dm)$. Then at the level of precision considered here,
    $\DM^2$ is proportional to $\dm$. So to extract $\dm$ one needs to know the
    physical value of $\DM^2$ and the constant $B_2$.
    
    $B_2$ was recently computed in \citep{Durr:2013goa}, using the same QCD
    simulations as the ones considered in the present paper. To determine
    $\DM^2$, we consider the leading isospin expansion of the kaon mass
    splitting $\DM_K^2=M_{K^+}^2-M_{K_0}^2$:
    \begin{equation}
        \DM_K^2=C_K\alpha+D_K\DM^2
		    \label{eq:dmsqkexp}
    \end{equation}
    Results for $\DM_K^2$ obtained for different values of $\alpha$ and $\delta
    m$ from lattice QCD and QED simulations can be fitted to this expression to
    obtain the coefficients $C_K$ and $D_K$ and subsequently the value of
    $\DM^2$ corresponding to physical quark masses, from the experimental value
    of $\DM_K^2$.
    
    \prlsec{Summary of the lattice methodology} The lattice setup used for this
    project is very similar to the one already described in
    \citep{Borsanyi:2013lga}. The work is based on our set of lattice QCD
    simulations presented in \citep{Durr:2010aw}. It is composed of $47$
    $N_f=2+1$ QCD ensembles with pion masses down to $\unit{120}{\mega\eV}$,
    $5$ lattice spacings down to $0.054\,\fm$ and $16$ different volumes up to
    $(\unit{6}{\fm})^3$. These simulations were performed using a tree-level
    $\bigo(a)$-improved Wilson fermion action with $2$ steps of HEX smearing.
    For each QCD configuration, a QED one is generated using the non-compact
    Maxwell action in Coulomb gauge with the four-momentum zero mode fixed to
    $0$. The resulting $SU(3)\times U(1)$ configuration is then included in the
    Wilson-Dirac operator used to compute the valence quark propagators, with
    the appropriate electric charge. The valence light quark masses are tuned
    to explore the region where $\dm$ varies between $0$ and its physical
    value. For most QCD ensembles, the unit of charge for valence quarks is set
    to its physical value. On one particular QCD ensemble, we perform three
    valence analyses: two with close to physical $\delta m$ and a value of
    $\alpha$ either about twice or one-fourth its physical value, and a third
    with $\alpha\simeq 0$ and $\delta m\simeq 0$. A plot of the values of
    $M_{\bar dd}^2$ versus $M_{\bar uu}^2$ used in our valence datasets can be
    found in \cite[Fig. 1]{Borsanyi:2013lga}.
    
    In this setup, two approximations are made: the sea $u$ and $d$ quark
    masses have the same mass and they carry no electric charge (QED is
    quenched). It is straightforward to show that the splitting of the sea
    light-quark masses only affects isospin splittings at orders in the isospin
    expansion which are beyond those considered. Regarding the quenching of
    QED, large $N_c$ counting and $\SU(3)$ flavor symmetry suggests that the
    sea QED effects may represent $\bigo(10\%)$ of the $\bigo(\alpha)$
    contribution to a given isospin splitting \citep{Borsanyi:2013lga}.
    Considering the EM part of the kaon splitting, which is of particular
    interest here, the next-to-leading order (NLO) PQ$\chi$PT+QED calculation
    of \citep{Bijnens:2006mk} can be used to estimate the QED quenching
    effects. In \citep{Portelli:2012pn} we argue that they may represent 5\% of
    the $\bigo(\alpha)$ correction. Nevertheless, for giving the reader an idea
    of how such a quenching uncertainty may propagate to the other quantities
    studied in this paper, we retain the more conservative 10\% quenching
    uncertainty on $\DQEDM_K^2 = \alpha C_K$.
    
    \prlsec{The EM contribution to the kaon splitting}
    \begin{figure}[!t]
    \includegraphics[width=0.99\columnwidth]{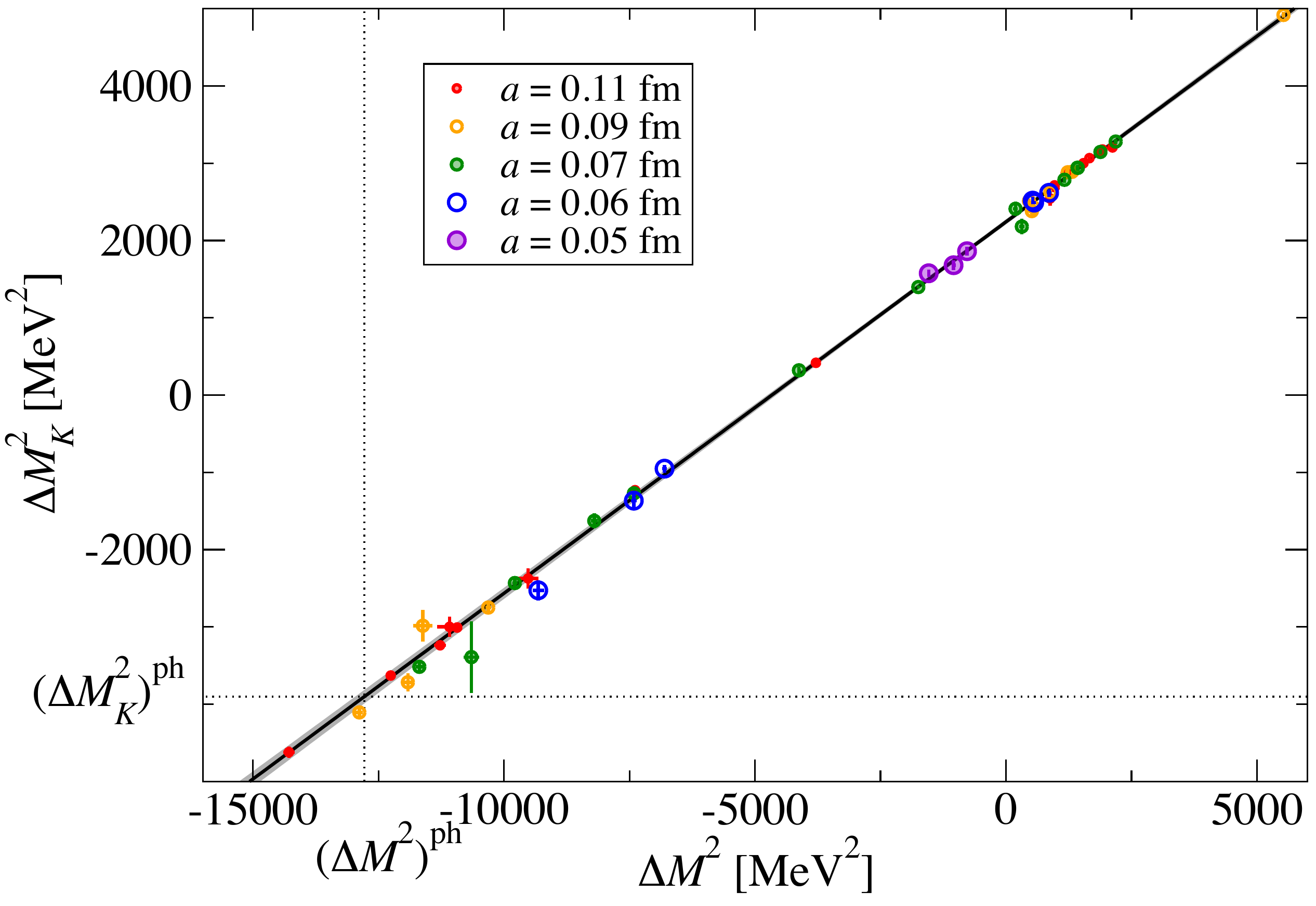}%
    \caption{\label{fig:DMKdm}Example of a fit of the dependence of
      $\DM_K^2$ on $\alpha$ and $\Delta M^2$ to the expression of
      Eq.~(\ref{eq:dmsqkexp}). Here, $\DM_K^2$ is plotted as a function of
      $\DM^2$. The dependence of the lattice results on all other
      variables has been subtracted using the fit. The fit has a
      correlated $\chi^2/\mathrm{dof}$ equal to $1.59$. It is plotted as a
      solid curve, with its 1$\sigma$ band.}
    \end{figure}
    \begin{figure}[!t]
    \includegraphics[width=0.99\columnwidth]{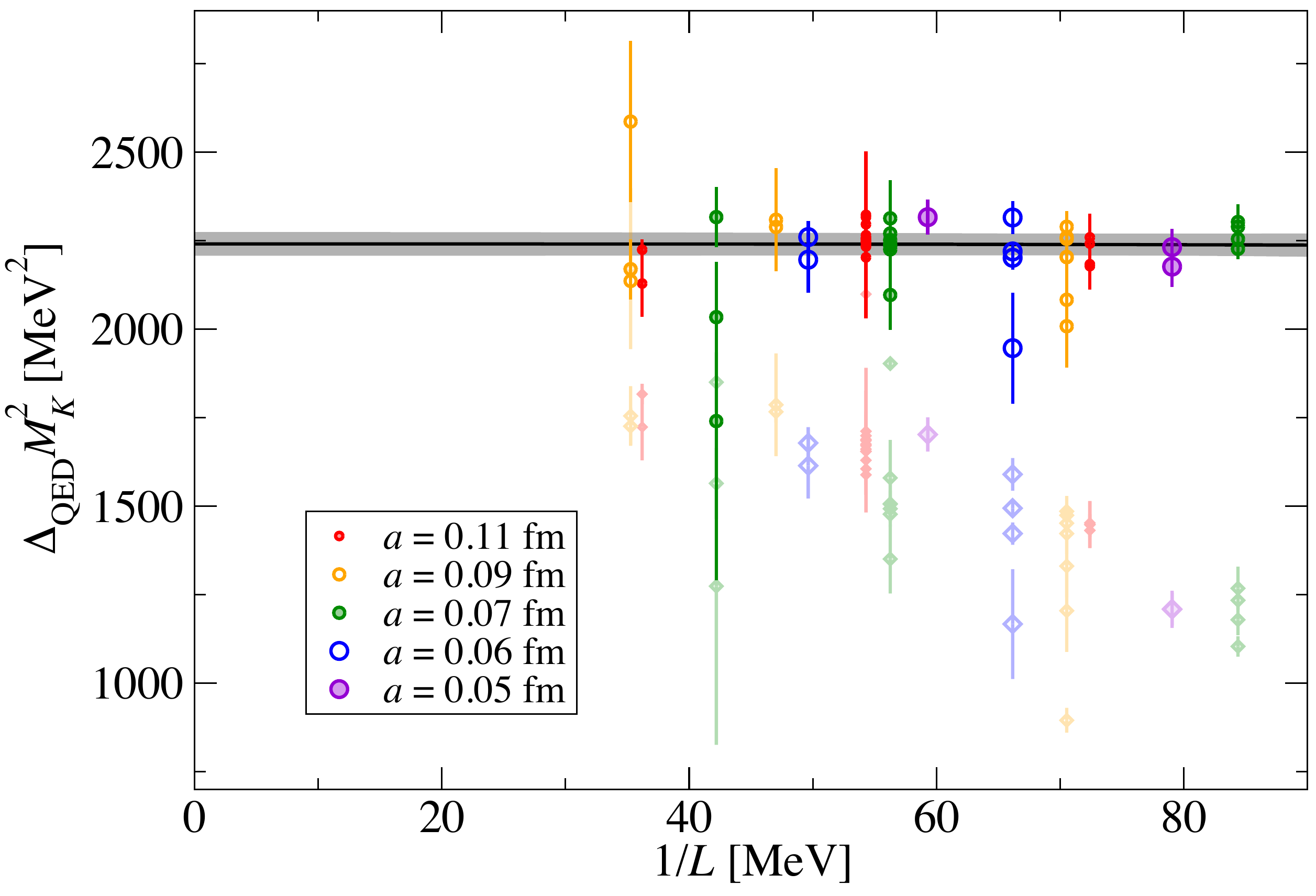}%
    \caption{\label{fig:DMKFV}Same fit as in Fig.~\ref{fig:DMKdm}. Here
    $\DQEDM_K^2$ is plotted as a function of $1/L$. The greyed symbols show the
    full volume dependence of the data. For the plain symbols, the universal
    $\bigo(1/L)$ and $\bigo(1/L^2)$ finite volume effects from~\cref{eq:fv}
    have been subtracted and the fit to the $\bigo(1/L^3)$ correction is
    plotted.}
    \end{figure}
    \begin{figure}[!t]
    \includegraphics[width=0.99\columnwidth]{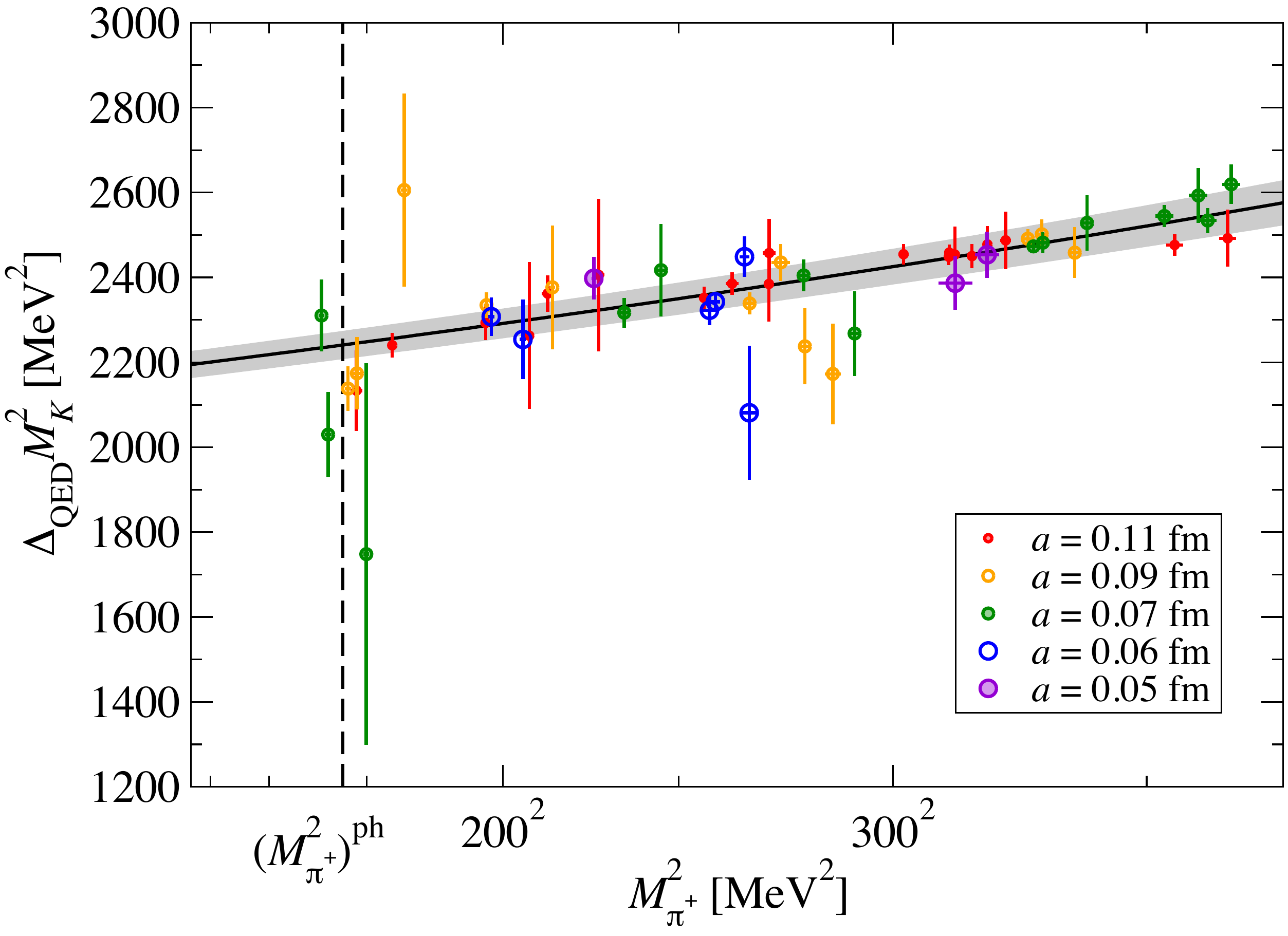}%
    \caption{\label{fig:DMKud}Same fit as in
      Fig.~\ref{fig:DMKdm}. Here $\DQEDM_K^2$ is plotted as a function of
      $M_{\pi^+}^2$.}
    \end{figure}
    In the expansion \cref{eq:dmsqkexp}, the coefficients $C_K$ and
    $D_K$ still depends on $m_{ud}$, $m_s$, $a$, and the temporal and spatial
    extents $T$ and $L$. We fix $m_{ud}$ and $m_s$ to their physical values by
    matching to the experimental values of $M_{\pi^+}^2$ and the combination
    $(M_{K^+}^2+M_{K^0}^2-M_{\pi^+}^2)/2$. Then, as explained in detail in
    \citep{Borsanyi:2013lga}, we use as a model for $\DM^2_K$ a first order
    expansion of $C_K$ and $D_K$ in these mass parameters around the physical
    mass point. Additionally, we allow for $\bigo(a)$ discretization effects
    and power-like $\bigo(1/L)$ finite-volume
    effects~\citep{Borsanyi:2014jba,Davoudi:2014qua} in the QED contribution
    proportional to $C_K$, as required in our setup. The finite volume effects
    are taken into account by adding the following term to the aforementioned
    expansion of $C_K$:
    \begin{equation}
      \label{eq:fv}
      C_K^{\mathrm{FV}}=-\frac{\kappa M_K}{L}\left[1+\frac{2}{M_KL}
      \left(1-\frac{\pi}{2\kappa}\frac{T}{L}\right)\right]+\frac{\rho}{L^3}
    \end{equation} 
    where $\kappa=2.837\ldots$ is a known number,
    $M_K=\frac12(M_{K^+}+M_{K^0})$ is the isospin averaged kaon mass and $\rho$
    is a fit parameter. In the parametrization~\cref{eq:fv}, the $\bigo(1/L)$
    and $\bigo(1/L^2)$ coefficients have been fixed to their known universal
    value~\citep{Borsanyi:2014jba} and the $\bigo(1/L^3)$ term is fitted to
    take into account additional structure-dependent effects. For the
    QCD contribution proportional to $D_K$, we assume $\bigo(\alpha_sa,a^2)$
    discretization effects and negligible finite-volume effects, which is
    justified in our large volumes given our present precision. To estimate
    systematic uncertainties, we consider a variety of analysis procedures.
    These variations are identical to those performed in
    \citep{Borsanyi:2013lga}. They include (please see \citep{Borsanyi:2013lga}
    for justifications and additional details): fitting the needed correlators
    on a conservative or a more aggressive time range; setting the scale with
    the mass of the $\Omega^-$ or the isospin-averaged $\Xi$; eliminating
    points with $M_{\pi^+}$ either greater than $\unit{400}{\mega\eV}$ or than
    $\unit{450}{\mega\eV}$ for the $\Omega^-$ and the $\Xi$ mass, and greater
    than $\unit{350}{\mega\eV}$ or than $\unit{400}{\mega\eV}$ for $\Delta
    M_K^2$; including either $\alpha_s a$ or $a^2$ contributions in $D_K$;
    replacing individually the Taylor mass expansions in $C_K$ and $D_K$ by the
    inverse of these expansions (for a total of $4$ choices). This leads to
    $128$ different determinations of $C_K$ and $D_K$. An example of such a fit
    is illustrated in \cref{fig:DMKdm,fig:DMKFV,fig:DMKud}. Finally, using the
    histogram method developed in \citep{Durr:2008zz}, we combine all of these
    results to obtain:
    \begin{equation}
      \label{eq:dqedmsqk}
      \DQEDM_K^2 = \unit{2186(26)(68)(219)}{\mega\eV\squared}
    \end{equation}
    where $C_K$ is taken at the physical mass point, in the continuum and
    infinite volume limits. Here the first error is statistical, the second is
    systematic, and the third is an estimation of the quenching uncertainty as
    discussed above. Our result can be compared to an estimate obtained from
    the input of FLAG \citep{Aoki:2013ldr},
    $\DQEDM_K^2=\unit{2090(380)}{\mega\eV\squared}$. The results are entirely
    compatible and ours has a total precision which is more than $5$ times
    higher omitting the generous estimate for the quenching error and more than
    $1.6$ times including it. For completeness, we also give the value of the
    slope of $\Delta M_K^2$ in $\Delta M^2$ at the physical point, obtained
    from our analysis: $D_K=0.484(5)(4)$. This result is compatible with the
    value $D_K=0.45(9)$, obtained by appropriately combining results from FLAG
    \citep{Aoki:2013ldr}. Its total error is $15$ times more precise.
    
    \prlsec{Corrections to Dashen's theorem}
	  As defined in \citep{Aoki:2013ldr}, one can quantify corrections to
    Dashen's theorem with the parameter:
    \begin{equation}
        \epsilon=\frac{\DQEDM_K^2-\DQEDM_{\pi}^2}{\DM_{\pi}^2}
    \end{equation}
    The pion isospin mass splitting, needed to evaluate $\epsilon$, is
    challenging to obtain through a lattice computation. Because the neutral
    pion is diagonal in flavor, correlation functions for this state will
    contain quark disconnected diagrams. These diagrams are known to be
    expensive and hard to evaluate on the lattice. Thus, we choose not to
    compute the pion splitting here. Fortunately, using $\mathrm{G}$-parity,
    one can easily show that the leading $\bigo(\dm)$ corrections to
    $\DM_{\pi}^2$ vanishes. Therefore, at the level of precision considered in
    this paper, we have $\DQEDM_{\pi}^2=\DM_{\pi}^2$, which is very well known
    experimentally \citep{Agashe:2014kda}.

    Using our result \cref{eq:dqedmsqk} for $\DQEDM_K^2$ and the experimental
    value of $\DM_{\pi}^2$, we obtain:
    \begin{equation}
      \epsilon=0.73(2)(5)(17)
      \label{eq:epsres}
    \end{equation}
    Here the relative quenching error, obtained by propagating a 10\%
    uncertainty in $\DQEDM_K^2$, is 23\%. Now, if we include an estimate of the
    $\delta m^2$ corrections in the relation of $\DQEDM_{\pi}^2$ to
    $\DM_{\pi}^2$, as given in \citep{Aoki:2013ldr} with the parameter
    $\epsilon_m=\DQCDM_\pi^2/\DM_{\pi}^2=0.04(2)$, we find $\epsilon =
    0.77(2)(5)(17)(2)$, with the fourth uncertainty coming from the one in
    $\epsilon_m$. Our result of (\ref{eq:epsres}) can be compared to the FLAG
    estimate $\epsilon=0.7(3)$ \citep{Aoki:2013ldr}. Again, it is entirely
    compatible with this estimate, and has a total precision which is more than
    $5$ times higher without the quenching uncertainty estimate and about $1.8$
    times higher with it.

    \prlsec{Up and down quark masses} 
    Using our analysis of the kaon splitting, the experimental value of this
    splitting, our lattice result $B_2=2.61(6)(1)\,\giga\eV$
    \citep{Durr:2013goa} in the $\bar{\mathrm{MS}}$ scheme at
    $\unit{2}{\giga\eV}$ and formula \cref{eq:dmsqkexp}, we obtain:
    \begin{equation}
        \label{eq:dm}
        \dm = m_u-m_d=\unit{-2.41(6)(4)(9)}{\mega\eV}
    \end{equation}
    in the same scheme and at the same scale. If one assumes a 10\% QED
    quenching error on $\DQEDM_K^2$, this error propagates to 3.7\% on $\dm$.
    It is interesting to note that the quenching of QED has a rather small
    impact on the determination of $\dm$. This comes essentially from the fact
    that the QCD part of the kaon splitting is roughly $3$ times larger than
    the QED part. Our result is entirely compatible with the value $\dm =
    \unit{-2.53(16)}{\mega\eV}$, derived from FLAG input \citep{Aoki:2013ldr}.
    Its precision is a factor of $2.1$ to $1.3$ higher, depending on
    whether the quenching error is taken into account.
    
    If we combine \cref{eq:dm} with our previous result
    $m_{ud}=3.469(47)(48)\,\mega\eV$ \citep{Durr:2010vn}, we get:
    \begin{align}
      \label{eq:mures}
        m_u &= m_{ud} + \frac{\dm}{2} = \unit{2.27(6)(5)(4)}{\mega\eV}\\
      \label{eq:mdres}
        m_d &= m_{ud} - \frac{\dm}{2} = \unit{4.67(6)(5)(4)}{\mega\eV}
    \end{align}
    still in the $\bar{\mathrm{MS}}$ scheme at $\unit{2}{\giga\eV}$. With the
    same assumptions as before, the QED quenching error on the individual quark
    masses is estimated to be 1.8 and 0.9\%, respectively. Our results are
    nicely compatible with the FLAG values $m_u=\unit{2.16(9)(7)}{\mega\eV}$
    and $m_d=\unit{4.68(14)(7)}{\mega\eV}$.
    
    From the results of (\ref{eq:mures}) and (\ref{eq:mdres}), we obtain the
    ratio of light quark masses:
    \begin{equation}
        \frac{m_u}{m_d} = 0.485(11)(8)(14)
    \end{equation}
    Strictly speaking, because $u$ and $d$ have different electric charges,
    this ratio is scale dependent in QCD plus QED. However it is easy to see
    that this dependency is beyond the leading isospin order considered in this
    work. Error propagations give a 2.9\% QED quenching uncertainty on this
    ratio. Our result is compatible with the FLAG average $m_u/m_d=0.46(2)(2)$.
    Moreover, our total precision is between $1.5$ and $2.2$ times higher,
    depending on whether our estimate of quenching uncertainties is included.
    
    We can further use our previous result $m_s/m_{ud}=27.53(20)(8)$
    \citep{Durr:2010vn} to build the flavor breaking ratios $R$ and $Q$:
    \begin{align}
        R&=\frac{m_s-m_{ud}}{m_d-m_u}=38.2(1.1)(0.8)(1.4)\\
        Q&=\sqrt{\frac{m_s^2-m_{ud}^2}{m_d^2-m_u^2}}=23.4(0.4)(0.3)(0.4)
    \end{align}
    QED quenching effects of order 4\% and 2\%, respectively, cannot be
    excluded on these quantities. It is
    also interesting to compare our results for $R$ to those obtained from
    $\chi$PT applied to $\eta\to 3\pi$ decays
    \citep{Gasser:1984pr,Bijnens:2007pr,Kambor:1995yc,Kampf:2011wr,Guo:2015zqa}.
    The convergence of $\chi$PT for this process is very poor and it is usually
    supplemented by a dispersive analysis. Without such an analysis, the
    results vary from 19.1 at LO to 31.8 at NLO and 42.2 (or 38.7 setting the
    $\bigo(p^6)$ low-energy constants to 0) at NNLO~\citep{Bijnens:2007pr}. The
    most recent NNLO $\chi$PT dispersive analysis~\citep{Kampf:2011wr} gives
    $R=37.7(2.2)$, in good agreement with our result.

    To summarize, our results are compatible with the estimates of
    \citep{Aoki:2013ldr}, which already include input from the quenched QED
    studies mentioned above~\footnote{In fact, preliminary versions of the
    results presented here
    \citep{Portelli:2010yn,Portelli:2012pn,Portelli:2015wna} impacted the
    analysis of \citep{Aoki:2013ldr}.}. In most cases, they significantly
    improve on their precision. In all isospin symmetry breaking quantities the
    quenching uncertainty is the dominant one. Therefore, it is now important
    to determine these quantities using a fully unquenched calculation with
    significantly higher statistics, such as the one carried out in
    \citep{Borsanyi:2014jba} for the splitting of stable hadrons.
    
    \medskip
    \begin{acknowledgments}
    L.L. thanks Heiri Leutwyler and A.P. thanks Émilie Passemar for
    enlightening discussions. C.H., L.L. and A.P. acknowledge the warm welcome
    of the Benasque Center for Science where this paper was partly completed.
    Computations were performed using the PRACE Research Infrastructure
    resource JUGENE and JUROPA based in Germany at FZ J\"ulich, as well as HPC
    resources provided by the ``Grand équipement national de calcul intensif''
    (GENCI) at the ``Institut du développement et des ressources en informatique
    scientifique'' (IDRIS) (GENCI-IDRIS Grant No. 52275), as well as clusters
    at Wuppertal and CPT. This work was supported in part by the OCEVU Labex
    (ANR-11-LABX-0060) and the A*MIDEX project (ANR-11-IDEX-0001-02), by CNRS
    grants GDR $n^0$2921 and PICS $n^0$4707, by EU grants FP7/2007-2013/ERC
    $n^0$208740, MRTN-CT-2006-035482 (FLAVIAnet), by DFG grants FO 502/2, SFB
    TRR-55 and UK STFC Grants ST/L000296/1 and ST/L000458/1. L.V. was partially supported by a GSI grant.
    \end{acknowledgments}

    \bibliography{udqmasses}

\end{document}